
e the fermionic
degrees of freedom of the gravitino field
are very restricted,     we have found two bosonic quantum
physical states, namely the wormhole  and the Hartle-Hawking state.
 From  the point of view of perturbation theory, it seems that  the
gravitational and gravitino modes that are allowed to be excited in a
supersymmetric Bianchi-IX model contribute in such a way to
forbid any physical solutions of the quantum constraints.
This suggests that in a complete perturbation expansion we would have to
conclude that the
full theory of N=1 supergravity with a non-zero cosmological constant
should have no physical states.
\\
mplete perturbation expansion we would have to conclude that the
full theory of N=1 supergravity with a non-zero cosmological constant
should have no physical states.
\\

\magnification\magstep1

\openup .5\jot

\input mssymb
\def\hbar{\mathchar '26\mkern -9muh}

\catcode`@=11
\def\eqaltxt#1{\displ@y \tabskip 0pt
  \halign to\displaywidth {%
    \rlap{$##$}\tabskip\centering
    &\hfil$\@lign\displaystyle{##}$\tabskip\z@skip
    &$\@lign\displaystyle{{}##}$\hfil\tabskip\centering
    &\llap{$\@lign##$}\tabskip\z@skip\crcr
    #1\crcr}}
\def\eqallft#1{\displ@y \tabskip 0pt
  \halign to\displaywidth {%
    $\@lign\displaystyle {##}$\tabskip\z@skip
    &$\@lign\displaystyle{{}##}$\hfil\crcr
    #1\crcr}}
\catcode`@=12 

\def\half{{\textstyle {1 \over 2}}}

\def\pmb#1{\setbox0=\hbox{#1}  \kern-.025em\copy0\kern-\wd0
  \kern0.05em\copy0\kern-\wd0  \kern-.025em\raise.0433em\box0 }
\def\pmbh#1{\setbox0=\hbox{#1} \kern-.12em\copy0\kern-\wd0
            \kern.12em\copy0\kern-\wd0\box0}
\def\sqr#1#2{{\vcenter{\vbox{\hrule height.#2pt
      \hbox{\vrule width.#2pt height#1pt \kern#1pt
         \vrule width.#2pt}
      \hrule height.#2pt}}}}

\def\rchi{{\raise 2pt \hbox {$\chi$}}}
\def\rga{{\raise 2pt \hbox {$\gamma$}}}
\def\rg{{\raise 2 pt \hbox {$g$}}}

\def\susy{supersymmetry}
\def\({\left(}
\def\){\right)}
\def\<{\left\langle}
\def\>{\right\rangle}

\def\[{\left[}
\def\]{\right]}
\let\text=\hbox
\def\pt{\partial}
\def\eps{\epsilon}
\def\kap{\kappa}

\def\om{\omega}
\def\ol{\overline}

\def\de{\delta}
\def\lam{\lambda}

\def\sig{\sigma}

\def\ti{\tilde}

\def\cH{{\cal H}}

\def\Lam{\Lambda}

\def\wti{\widetilde}

\hfuzz 6pt

\catcode`@=12 
\rightline {\bf DAMTP R94/20}
\rightline {\bf REVISED VERSION}
\centerline {\bf Quantization of  Bianchi Models in N=1 Supergravity}
\centerline {\bf with a Cosmological Constant\footnote{$\dagger$}{{\rm Lecture
presented at the International School-Seminar
"Multidimensional Gravity and Cosmology", Yaroslavl,
 Russia, June 20 -- 26, 1994; to be published by World Scientific, ed. V.
Melnikov}}}
\vskip .1 true in
\centerline {A.D.Y. Cheng, P.D. D'Eath and
$\underline{{\rm P.R.L.V. Moniz}}$\footnote*{{\rm e-mail address:
prlvm10@amtp.cam.ac.uk}}}
\vskip .1 true in
\centerline {Department of Applied Mathematics and Theoretical Physics}
\centerline {University of Cambridge, Silver Street,Cambridge CB3 9EW, UK }
\vskip .2 true in
\centerline {\bf ABSTRACT}
\vskip .1 true in

{\sevenrm We study the quantization of some cosmological  models within the
theory of N=1 supergravity with
a positive cosmological constant. We find, by imposing the supersymmetry and
Lorentz constraints, that there
are {\seveni no} physical states in the models we have considered.
For  the  k=1 Friedmann-Robertson-Walker model, where the fermionic
degrees of freedom of the gravitino field
are very restricted,     we have found two bosonic quantum
physical states, namely the wormhole  and the Hartle-Hawking state.
 From  the point of view of perturbation theory, it seems that  the
gravitational and gravitino modes that are allowed to be excited in a
supersymmetric Bianchi-IX model
contribute in such a way to
forbid any physical solutions of the quantum constraints.
This suggests that in a complete perturbation expansion we would have to
conclude that the
{\seveni full} theory of N=1 supergravity with a non-zero cosmological constant
should have {\seveni no} physical states.}

\vfill
\magnification\magstep1
\centerline {PACs numbers: 04.60.+ $n$, 04.65.+ $e$, 98.80. $Hw$ }
\vskip 6 pt
\noindent

\noindent

\magnification\magstep1

\openup .5\jot

\input mssymb
\def\hbar{\mathchar '26\mkern -9muh}

\catcode`@=11
\def\eqaltxt#1{\displ@y \tabskip 0pt
  \halign to\displaywidth {%
    \rlap{$##$}\tabskip\centering
    &\hfil$\@lign\displaystyle{##}$\tabskip\z@skip
    &$\@lign\displaystyle{{}##}$\hfil\tabskip\centering
    &\llap{$\@lign##$}\tabskip\z@skip\crcr
    #1\crcr}}
\def\eqallft#1{\displ@y \tabskip 0pt
  \halign to\displaywidth {%
    $\@lign\displaystyle {##}$\tabskip\z@skip
    &$\@lign\displaystyle{{}##}$\hfil\crcr
    #1\crcr}}
\catcode`@=12 

\def\half{{\textstyle {1 \over 2}}}

\def\pmb#1{\setbox0=\hbox{#1}  \kern-.025em\copy0\kern-\wd0
  \kern0.05em\copy0\kern-\wd0  \kern-.025em\raise.0433em\box0 }

\def\pmbh#1{\setbox0=\hbox{#1} \kern-.12em\copy0\kern-\wd0
            \kern.12em\copy0\kern-\wd0\box0}
\def\sqr#1#2{{\vcenter{\vbox{\hrule height.#2pt
      \hbox{\vrule width.#2pt height#1pt \kern#1pt
         \vrule width.#2pt}
      \hrule height.#2pt}}}}

\def\rchi{{\raise 2pt \hbox {$\chi$}}}
\def\rga{{\raise 2pt \hbox {$\gamma$}}}
\def\rrho{{\raise 2pt \hbox {$\rho$}}}

\def\({\left(}
\def\){\right)}
\def\<{\left\langle}
\def\>{\right\rangle}

\def\[{\left[}
\def\]{\right]}
\let\text=\hbox
\def\pt{\partial}
\def\eps{\epsilon}
\def\kap{\kappa}

\def\om{\omega}
\def\ol{\overline}

\def\de{\delta}
\def\lam{\lambda}

\def\sig{\sigma}

\def\ti{\tilde}

\def\Lam{\Lambda}

\def\wti{\widetilde}

\def\lrta{\longrightarrow}
\def\cH{{\cal H}}


{\bf I. Introduction}

Recently a number of quantum cosmological models have been studied in which
the action is that of supergravity, with possible additional coupling to
supermatter [1-11,14,15,18,26-29,31-34,38,41].
 In addition, a review on this rather fascinating subject is under preparation
[12]. It is sufficient, in finding a physical state, to solve
the Lorentz and supersymmetry constraints of the theory [13,14]. Because of
the anti-commutation relations $ \[ S_A,~\wti S_{A'} \]_+ \sim \cH_{A A'} $,
the supersymmetry constraints $ S_A \Psi = 0,~\ol S_{A'} \Psi = 0 $ on a
physical wave function $ \Psi $ imply the Hamiltonian constraint $ \cH_{A A'}
\Psi = 0 $ [13,14].

In the case of the Bianchi-I model in $ N = 1 $ supergravity with no
cosmological constant ($ \Lam = 0 $) [8], the quantum states are
in the bosonic and filled fermionic sectors and are of the form
$\exp (-\half h^{-\half})$, where $ h = \det h_{i j} $ is the
determinant of the three-metric. In the case of Bianchi IX with $ \Lam = 0
$, there are
 two states, of the form $ \exp ( \pm I / \hbar ) $ where $ I $ is a
certain Euclidean action, one in the empty and one in the filled fermionic
sector [9,15].
When the usual choice of spinors constant in the standard basis is made
for the gravitino field, the bosonic state $ \exp ( - I / \hbar ) $ is the
wormhole state [9,16]. With a different choice, one obtains the
Hartle--Hawking state [15,17]. Similar states were found for $ N = 1 $
supergravity in the more general Bianchi models of class $ A $ [10].
[Supersymmetry (as well as other considerations) forbids mini-superspace
models of class $ B $.]

It is of interest to extend these results, by studying more general locally
supersymmetric actions, initially in Bianchi models. Possibly the simplest
such generalization is the addition of a cosmological constant in $ N = 1 $
supergravity [19]. On the one hand, the appearance of a cosmological constant
term in some supergravity models is a consequence of the coupling to matter.
In particular, when one gauges internal SO(2) or SO(3) symmetries in
N=2, N=3 {\it extended} supergravities [20]  (coupling the spin-1 fields in a
electromagnetic way to the gravitino) one needs at the same time a
cosmological constant and a mass-like term for the gravitino in the action
[20--24]. Such interesting connection between spin-1 fields
and a cosmological constant has led to the suggestion that the electromagnetism
might be  due to a De Sitter space-time curvature within a supergravity
context [20]. Furthermore, the action for our model presented in eq. (2.1)
can be obtained from  the O(2)-gauge extended supergravity model
when one eliminates from the lagrangian the spin ($ {3 \over 2} $,1)-multiplet
while
keeping a non-zero gauge coupling constant [25]. But, on the other hand,
our model can also be derived as an extension of pure N=1
supergravity [19]. Further, little has been written about extensions of
pure supergravity to include $R^2$ terms, etc. The main idea in this extension
is based on the fact that as the presence of a non zero cosmological constant
induces a constant curvature of space-time independent of matter, the symmetry
properties of such spaces will be in correspondence with the De Sitter group
rather than  the Poincar\'e group. The new action is determined by the
prescription that quantities such as the covariant derivative and curvature
terms which are characteristic of the Poincar\'e group should be replaced
in the {\it field equations} by new ones which are charactristic of the De
Sitter
group. Following this procedure, one concludes that even in
N=1 supergravity one needs a mass-like term for the spin-$ {3 \over 2} $ field
if
one adds a non-zero cosmological constant even though there are no
spin-1 fields.

Using the triad ADM
canonical formulation we
shall see that there are  no physical quantum states in
the cases of Bianchi type-I and IX (diagonal) models [26,27].  The
calculations are described in Sec.~II. We also treat briefly in Sec.~III the
spherical $ k = + 1 $ Friedmann model, and find that there is a two-parameter
family of solutions of the quantum constraints with a $ \Lam $-term.
Nevertheless, as will be seen, the Bianchi-IX model provides a better guide
to the generic result, since more spin-$ {3 \over 2} $ modes are available to
be excited in the Bianchi-IX model, while the form of the fermionic fields
needed for supersymmetry in the $ k =
+ 1 $ Friedmann model is very restrictive [6]. In Sec. IV we briefly comment
on how other different approaches from the one presented in the
previous sections allows one to extract some similar results [4,5,28,29],
[30-34].
Sec.~V contains the
Conclusion.
\medbreak
\noindent
{\bf II. QUANTUM STATES FOR THE BIANCHI MODELS WITH A $ \Lam $-TERM}

Using two-component spinors [6,14], the action [19] is
$$ S = \int d^4 x \[ \eqalign {
{}~& \( 2 \kap^2 \)^{- 1} \( \det e \)~\( R - 3 g^2 \) \cr
+ &{1 \over 2} \eps^{\mu \nu \rho \sig} \( \ol \psi^{A'}_{~~\mu} e_{A A' \nu}
D_\rho \psi^A_{~~\sig} + H.c. \) \cr
- &{1 \over 2} g (\det e)~\( \psi^A_{~~\mu} e_{A B'}^{~~~~\mu} e_B^{~~B' \nu}
\psi^B_{~~\nu} + H.c. \) \cr }
\]~. \eqno (2.1) $$
Here the tetrad is $ e^a_{~\mu} $ or equivalently $ e^{A A'}_{~~~~\mu} $. The
gravitino field $ \(\psi^A_{~~\mu}, \ol \psi^{A'}_{~~\mu} \) $ is
an odd (anti-commuting) Grassmann quantity. The scalar
curvature $ R $ and the covariant derivative $ D_\rho $ include torsion.
We define $ \kap^2 =
8 \pi $. Here $ g $ is a constant, and the cosmological constant is $ \Lam =
{3 \over 2} g^2 $.

There are two possible approaches to the quantization of this model. One
possibility is to substitute the Bianchi Ansatz, e.g.,
$$ d s^2 = \( N^j N_j - N^2 \) d t^2 + 2 N_i d t~d x^i + h_{i j} d x^i d
x^j~, \eqno (2.2) $$
for the geometry $ e^{A
A'}_{~~~~\mu} $, where all metric components are functions of time,
 and gravitino field $ \( \psi^A_{~~\mu}, \ol
\psi^{A'}_{~~\mu} \) $ into the action (2.1). The components $ \psi^A_{~~\mu}
e^{B B' \mu} $ and $ \ol \psi^{A'}_{~~\mu} e^{B B' \mu} $ are required to be
spatially constant with respect to the standard triad [35] on the Bianchi
 hypersurfaces. One finds that, in order for the form of the
Ansatz to be left
invariant by one-dimensional local supersymmetry transformations, possibly
corrected by coordinate and Lorentz transformations [6], one must study the
general non-diagonal Bianchi model [35]. The reduced action could then be
computed, leading to the Hamiltonian standard form
$$ H = \wti N \cH + \rrho_A S^A + \wti S^{A'} \ti \rrho_{A'} + M_{A B} J^{A B}
+ \wti M_{A' B'} \wti J^{A' B'}~. \eqno (2.3) $$
Hence $ \cH $ is the generator of local time translations, $ S^A $ and $
\wti S^{A'} $ are the generators of local supersymmetry transformations, and $
J^{A B} $ and $ \wti J^{A' B'} $ are the generators of local Lorentz
transformations; they are formed from the basic dynamical variables $ \( e^{A
A'}_{~~~~i} , \psi^A_{~~i} , \wti \psi^{A'}_{~~i} \) $. [At this point it is
natural in the classical theory to free $ \wti \psi^{A'}_{~~i} $ from being
the hermitian conjugate of $ \psi^A_{~~i} $.] The quantities $ \wti N,
\rrho_A, \ti \rrho_{A'}, M_{A B} $ and $ \wti M_{A' B'} $ are Lagrange
multipliers.
The supersymmetry and Lorentz constraints are imposed on physical wave
functions
but they would be complicated because of the number of parameters needed to
describe the off-diagonal model.

The other alternative, taken here, is to apply the supersymmetry constraints
of the general theory at a  Bianchi geometry [9]. This is valid
since the supersymmetry constraints are of first order in bosonic
derivatives, and give expressions such as $ \de \Psi / \de h_{i m} (x) $ in
terms of known quantities and $ \Psi $. These equations can be evaluated
in the case of,
e.g., a
diagonal Bianchi-IX geometry, parametrized by three radii $ A,~B,~C $. One
multiplies (e.g.) by $ \de h_{i m} (x) = \pt h_{i m} / \pt A $ and integrates $
\int d^3 x (~~) $ to obtain an equation for $ \pt \Psi / \pt A $ in terms of
known
quantities. The need to consider off-diagonal metrics is thereby avoided.

In general, it is only necessary to solve the quantum constraints
$$
S^A \Psi = 0~,\bar S^{A'} \Psi = 0~,
J^{AB} \Psi = 0~,  \bar J^{A' B'} \Psi = 0~, \eqno (2.4)   $$
for a physical state $ \Psi $, since the anti-commutator of $ S^A $ and $
\bar S^{A'} $ includes $ \cH^{AA'} $, so that Eq.~(2.4) implies also $
\cH^{AA'} \Psi = 0
$. The wave function can (e.g.) be taken as $ \Psi \( e^{A A'}_{~~~~i},
\psi^A_{~~i} \) $ or $ \wti \Psi \( e^{A A'}_{~~~~i}, \wti \psi^{A'}_{~~i} \)
$. These representations are related by a fermionic Fourier transform [5, 9].

Classically, the   supersymmetry constraints are
$$ S_A = g h^\half e_A^{~~A' i} n_{B A'} \psi^B_{~~i} +
\eps^{i j k} e_{A B' i}~^{3 s} D_j\ol \psi^{B'}_{~~k}
{}~- \half i \kap^2 p_{A A'}^{~~~~i} \wti \psi^{A'}_{~~i}
{}~, \eqno (2.5) $$
$$ \ol S_{A'} = g h^{1 \over 2} e^{A~~~i}_{~~A'} n_{A B'} \ol \psi^{B'}_{~~i}
+ \eps^{i j k} e_{A A' i}~^{3 s} D_j \psi^A_{~~k} + {1 \over 2} i \kap^2
\psi^A_{~~i} p_{A A'}^{~~~~i}~. \eqno (2.6) $$
 Here $ n^{A A'} $ is the spinor version of the
unit future-pointing normal $ n^\mu $ to the constant  $ t $ surface. It
is a function of the $ e^{A A'}_{~~~~i} $, defined by
$$ n^{A A'} e_{A A' i} = 0~, \ \ \ \ n^{A A'} n_{A A'} = 1~. \eqno (2.7) $$
In Eq.~(2.5),(2.6), $ p_{A A'}^{~~~~i} $ is the momentum conjugate to $ e^{A
A'}_{~~~~i} $. The expression $ ^{3 s}D_j $ denotes the three-dimensional
covariant derivative without torsion. Since the components of $ \psi^A_{~~k}
$ are taken to be constant in the Bianchi basis, one can replace $
^{3s}D_j \psi^A_{~~k} $ by $ \om^A_{~~B j} \psi^B_{~~k} $, where $ \om^A_{~~B
j} $ gives the torsion-free connection [14].

Quantum-mechanically, in the representation $ \Psi \( e^{A A'}_{~~~~i},
\psi^A_{~~i} \) $, one represents [9,14]
$$ p_{A A'}^{~~~~i} \lrta - i \hbar {\de \over \de e^{A A'}_{~~~~i}} + \half
\eps^{i j k} \psi_{A j} \bar \psi_{A' k}~, \eqno (2.8) $$
where
$$ \bar \psi^{A'}_{~~i} \lrta - i \hbar D^{A A'}_{~~~~j i}
 h^{1 \over 2}
{\pt \over \pt
\psi^A_{~~j}}~, \eqno (2.9) $$
where $ \pt / \pt \psi^A_{~~j} $ denotes left differentiation [13], and
$$ D^{A A'}_{~~~~j i} = - 2 i h^{- \half} e^{A B'}_{~~~~i} e_{B B' j} n^{B
A'}~. \eqno (2.10) $$
We have made
the replacement $ \de \Psi / \de \psi^B_{~~j} \longrightarrow h^{1 \over 2}
\pt \Psi / \pt \psi^B_{~~j} $.
This replacement is important when considering space-time manifolds
whose spatial sections are compact.
The $ h^{1 \over 2} $ factor ensures that each
term has the correct weight in the equations, namely when one takes a
variation of a (compact) Bianchi geometry, multiplying by
 $ \de  / \de h_{i j} $ and integrating over the
three-geometry (see eq. (2.21),(2.23), (2.24)).
One can  check, e.g., that this
replacement gives the correct supersymmetry constraints in the $ k = + 1 $
Friedmann model, where the model was quantized using
the alternative approach via a supersymmetric Ansatz [6].

The corresponding quantum constraints read, with the help of [14],
$$ \eqalignno {
\ol S_{A'} \Psi &= - i \hbar g h^{1 \over 2} e^{A~~~i}_{~~A'} n_{A B'} D^{B
B'}_{~~~~j i} \( h^{1 \over 2} {\pt \Psi \over \pt \psi^B_{~~j}} \) \cr
{}~&+ \eps^{i j  k} e_{A A' i} \om^A_{~~B j} \psi^B_{~~k} \Psi - {1 \over 2}
\hbar
\kap^2 \psi^A_{~~i} {\de \Psi \over \de e^{A A'}_{~~~~i}} = 0~, &(2.11) \cr
S_A \Psi &= g h^{1 \over 2} e_A^{~~A' i} n_{B A'} \psi^B_{~~i} \Psi - i \hbar
\om_{A~~i}^{~~B} \( h^{1 \over 2} {\pt \Psi \over \pt \psi^B_{~~i}} \) \cr
{}~&+ {1 \over 2} i \hbar^2 \kap^2 D^{B A'}_{~~~~j i} \( h^{1 \over 2} {\pt
\over
\pt \psi^B_{~~j}} \)~{\de \Psi \over \de e^{A A'}_{~~~~i}} = 0~. &(2.12) \cr
} $$

The constraints $ J^{A B} \Psi = 0,~~\bar J^{A' B'} \Psi = 0 $ imply that $
\Psi \( e^{A A'}_{~~~~i}, \psi^A_{~~i} \) $ is a Lorentz-invariant function.
One solves them by taking expressions in which all spinor indices have been
contracted together. As described in [5], it is reasonable also to consider
only wave functions $ \Psi $ which are spatial scalars, where all spatial
indices $ i, j, \ldots $ have also been contracted together.
 To specify this, note the
decomposition [11] of $ \psi^A_{~~B B'} = e_{B B'}^{~~~~i} \psi^A_{~~i} $:
$$ \psi_{A B B'} = - 2 n^C_{~~B'} \rga_{A B  C} + {2 \over 3} \( \beta_A
n_{B B'} + \beta_B n_{A B'} \) - 2 \eps_{A B} n^C_{~~B'} \beta_C~, \eqno
(2.13) $$
where $ \rga_{A B C} = \rga_{(A B C)} $ is totally symmetric and $
\eps_{A B} $ is the alternating spinor. The general Lorentz-invariant wave
function is a polynomial of sixth degree in Grassmann variables:
$$ \eqalignno {
\Psi \( e^{A A'}_{~~~~i},~\psi^A_{~~i} \) &= \Psi_0 \( h_{i j} \) + \( \beta_A
\beta^A \) \Psi_{21} \( h_{i j} \) + \( \rga_{A B C} \rga^{A B C} \)
\Psi_{22} \( h_{i j} \) \cr
{}~&+ \( \beta_A \beta^A \)~\( \rga_{B C D} \rga^{B C D} \) \Psi_{41} \( h_{i
j} \) + \( \rga_{A B C} \rga^{A B C} \)^2 \Psi_{42} \( h_{i j} \) \cr
{}~&+ \( \beta_A \beta^A \)~\( \rga_{B C D} \rga^{B C D} \)^2 \Psi_6 \( h_{i
j} \)~. &(2.14) \cr
} $$
Any other Lorentz-invariant fermionic polynomials can
be written in terms of these. Note that, for example,
the term $ \( \beta^A \rga_{A B C} \)^2 = \beta^A
\rga_{A B C} \beta^D \rga_D^{~~B C} $ can be rewritten, using the
anti-commutation of the $ \beta $'s and $ \rga $'s, as
$$ {\rm const.}~\beta^E \beta_E \eps^{A D} \rga_{A B C} \rga_D^{~~B C} = {\rm
const.}~ \( \beta_E \beta^E \)~\(\rga_{A B C} \rga^{A B C} \).~\eqno(2.15) $$
Similarly, any quartic in $ \rga_{A B C} $ can be rewritten as a multiple of
$ \( \rga_{A B C} \rga^{A B C} \)^2 $. Since there are only four independent
components of $ \rga_{A B C} = \rga_{(A B C)} $, only one independent quartic
can be made from $ \rga_{A B C} $, and it is sufficient to check that $ \(
\rga_{A B C} \rga^{A B C} \)^2 $ is non-zero. Now $ \rga_{A B C} \rga^{A B C}
= 2
\rga_{000} \rga_{111} - 6 \rga_{100} \rga_{011} $. Hence $ \( \rga_{A B C}
\rga^{A B C} \)^2 $ includes a non-zero quartic term const.~$\rga_{000}
\rga_{100} \rga_{110} \rga_{111} $.
Unlike the case of $ N = 1 $ supergravity [5], here the nonzero $ g $ (or $
\Lam $) implies that there is coupling between different fermionic levels.

We now proceed to solve the supersymmetry and Lorentz constraints for the
case
of a diagonal Bianchi-IX [29], whose three-metric is given in terms of the
three radii $
A, B, C $ by
$$ h_{i j} = A^2 E^1_{~i} E^1_{~j} + B^2 E^2_{~i} E^2_{~j} + C^2 E^3_{~i}
E^3_{~j}~, \eqno (2.16) $$
where $ E^1_{~i}, E^2_{~i}, E^3_{~i} $ are a basis of unit left-invariant
one-forms on the three-sphere [35]. In the calculation, we shall repeatedly
need the expression:
$$ \eqalignno {
\om_{A B i} n^A_{~~B'} e^{B B' j} &= {i \over 4} \( {C \over A B} + {B \over C
A} - {A \over B C} \) E^1_{~i} E^{1 j} \cr
{}~&+ {i \over 4} \( {A \over B C} + {C \over A B} - {B \over C A} \) E^2_{~i}
E^{2 j} \cr
{}~&+ {i \over 4} \( {B \over C A} + {A \over B C} - {C \over A B} \) E^3_{~i}
E^{3 j} &(2.17) \cr } $$
This can be derived from the expressions for $ \om^{A B}_{~~~i} $ given in
[9,13].

First consider the $ \ol S_{A'} \Psi = 0 $ constraint at the level $ \psi^1 $
in powers of fermions. One obtains
$$ {3 \over 16} \hbar g h^{1 \over 2} e_{B A'}^{~~~~i} \psi^B_{~~i} \Psi_{21}
+ \eps^{j k i} e_{A A' j} \om^A_{~~B k} \psi^B_{~~i} \Psi_0 + \hbar \kap^2
e_{B A' j} \psi^B_{~~i} {\de \Psi_0 \over \de h_{i j}} = 0~. \eqno (2.18) $$
Since this holds for all $ \psi^B_{~~i} $, one can conclude
$$ {3 \over 16} \hbar g h^{1 \over 2} e_{B A'}^{~~~~i} \Psi_{21} + \eps^{j k
i} e_{A A' j} \om^A_{~~B k} \Psi_0 + \hbar \kap^2 e_{B A' j} {\de \Psi_0
\over \de h_{i j}} = 0~. \eqno (2.19) $$
Now multiply this equation by $ e^{B A' m} $, giving
$$ - {3 \over 16} \hbar g h^{i m} h^{1 \over 2} \Psi_{21}
+ \eps^{j k i} e_{A A' j} e^{B A' m} \om^A_{~~B k} \Psi_0
- \hbar \kap^2 {\de \Psi_0 \over \de h_{i m}} = 0~. \eqno (2.20) $$
The second term can be simplified using [6]
$$ e_{A A' j} e^{B A'}_{~~~~m} = - { 1 \over 2} h_{j m} \eps_A^{~~B} + i
\eps_{j m n}
h^{1 \over 2} n_{A A'} e^{B A' n}~. \eqno (2.21) $$
One then notes, as above, that by taking a variation among the Bianchi-IX
metrics, such as
$$ \de h_{i j} = {\pt h_{i j} \over \pt A} = 2 A E^1_{~i} E^1_{~j}~, \eqno
(2.22) $$
multiplying by $ \de \Psi_0 / \de h_{i j} $ and integrating over the
three-geometry, one obtains $ \pt \Psi_0 / \pt A $. Putting this information
together one obtains the constraint
$$ \hbar \kap^2 {\pt \Psi_0 \over \pt A} + 1 6 \pi^2 A \Psi_0 + 6 \pi^2 \hbar
g B C \Psi_{21} = 0~, \eqno (2.23) $$
and two others given by cyclic permutation of $ A B C $.

Next we consider the $ S_A \Psi = 0 $ constraint at order $ \psi^1 $. One
uses the relations $ \pt \( \beta_A \beta^A \) / \pt \psi^B_{~~i} = -
n_A^{~~B'} e_{B B'}^{~~~~i} \beta^A $ and $ \pt \( \rga_{A D C} \rga^{A D C}
\) / \pt \psi^B_{~~i} = - 2 \rga_{B D C}~ n^{C C'} e^{D~~~~~i}_{~~~C'} $, and
writes out $ \beta^A $ and $ \rga_{B D C} $ in terms of $ e^{E E'}_{~~~~j} $
and $ \psi^E_{~~j} $. Proceeding by analogy with the previous calculation
above, one again `divides out' by $ \psi^B_{~~j} $.
One replaces the free spinor indices $ A B $ by the spatial index $ n $ on
multiplying by $ n^A_{~~D'} e^{B D' n} $, then
multiplying by different choices $ \de h_{i m} = \pt h_{i m} / \pt A $ etc.
and integrating over the manifold, one finds the constraints
$$ \eqalignno {
&{1 \over 16} \hbar^2 \kap^2 A^{- 1} \( A {\pt \Psi_{21} \over \pt A} + B
{\pt \Psi_{21} \over \pt B} + C {\pt \Psi_{21} \over \pt C} \) \cr
- &{1 \over 3} \hbar \kap^2 \[ 3 {\pt \Psi_{22} \over \pt A} - A^{- 1} \( A
{\pt \Psi_{22} \over \pt A} + B {\pt \Psi_{22} \over \pt B} + C {\pt
\Psi_{22} \over \pt C} \) \] \cr
&- 16 \pi^2 g B C \Psi_0
- \pi^2 \hbar B C \( {A \over B C} + {B \over C A} + {C \over A B} \)
\Psi_{21} \cr
&+ {1 \over 3} \( 16 \pi^2 \) \hbar B C \( {2 A \over B C} - {B \over C A} -
{C \over A B} \) \Psi_{22} = 0~. &(2.24) \cr
} $$
and two more equations given by cyclic permutation of $ A B C $.

Now consider the $ \ol S_{A'} \Psi = 0 $ constraint at order $ \psi^3 $. It
will turn out that we need go no further than this. The constraint can be
written as
$$ \eqalignno {
&{1 \over 2} \hbar g h^{1 \over 2} e^B_{~~A' j} n_C^{~~B'} e_{B B'}^{~~~~j}
\beta_C \( \rga_{D E F} \rga^{D E F} \) \Psi_{41} \cr
&+ \eps^{i j k} e_{A A' i} \om^A_{~~B j} \psi^B_{~~k} \[ \( \beta_C \beta^C
\) \Psi_{21} + \( \rga_{C D E} \rga^{C D E} \) \Psi_{22} \] \cr
&- {1 \over 2} \hbar^2 \kap^2 \psi^A_{~~i} \[ \( \beta_C \beta^C \) {\de
\Psi_{21} \over \de e^{A A'}_{~~~~i}} + \( \rga_{C D E} \rga^{C D E} \) {\de
\Psi_{22} \over \de e^{A A'}_{~~~~i}} \] = 0~. &(2.25) \cr
} $$
The terms $ \psi^B_{~~k} $ and $ \psi^A_{~~i} $ in the last two lines can be
rewritten in terms of $ \beta_A $ and $ \rga_{F G H} $, using Eq.~(2.13). Then
one can set separately to zero the coefficient of $ \beta^C \( \rga_{D E F}
\rga^{D E F}
\) $, the symmetrized coefficient of $ \rga_{D E F} \( \beta_C \beta^C \) $
and the symmetrized coefficient of $ \rga_{F G H} \( \rga_{C D E} \right .
$ \break $ \left . \rga^{C D E} \) $. These three equations give
$$ {3 \over 4} \hbar g h^{1 \over 2} n^C_{~~A'} \Psi_{41}
- {8 \over 3} \eps^{i j k} e_{A A' i} \om^A_{~~B j} n^B_{~~C'} e^{C
C'}_{~~~~k} \Psi_{22}
+ {4 \over 3} \hbar \kap^2 n^A_{~~B'} e^{C B'}_{~~~~i} {\de \Psi_{22} \over
\de e^{A A'}_{~~~~i}} = 0~, \eqno (2.26) $$
$$ 2 \eps^{i j k} e_{A A' i} \om^A_{~~B j} n^D_{~~B'} e^{C B'}_{~~~~k}
\Psi_{21}
- \hbar \kap^2 n^D_{~~B'} e^{C B'}_{~~~~i} {\de \Psi_{21} \over \de e^{B
A'}_{~~~~i}} $$
$$ + \( B C D \to C D B \) + \( B C D \to D B C \) = 0~, \eqno (2.27) $$
and Eq.~(2.27) with $ \Psi_{21} $ replaced by $ \Psi_{22} $. Contracting
Eq.~(2.26) with $ n_C^{~~A'} $ and integrating over the three-surface gives
$$ {3 \over 4} \( 16 \pi^2 \) \hbar g A B C \Psi_{41} + {2 \over 3} \( 16
\pi^2 \)~ \( A^2 + B^2 + C^2 \) \Psi_{22} $$
 $$ + {2 \over 3} \hbar \kap^2 \( A {\pt \Psi_{22} \over \pt A} + B {\pt
\Psi_{22} \over \pt B} + C {\pt \Psi_{22} \over \pt C} \) = 0~. \eqno (2.28)
$$
Contracting Eq.~(2.27) with $ e^{B A' \ell} n_{C C'} e_D^{~~C' N} $,
multiplying by $ \de h_{\ell n} = \pt h_{\ell n} / \pt A $ and integrating
gives
$$ 3 \hbar \kap^2 {\pt \Psi_{21} \over \pt A} - \hbar \kap^2 A^{- 1} \( A
{ \pt \Psi_{21} \over \pt A} + B {\pt \Psi_{21}
\over \pt B} + C {\pt \Psi_{21} \over \pt C} \) $$
$$ - 16 \pi^2 B C \( {C \over A B} + {B \over C A} - 2 {A \over B C} \)
\Psi_{21}
= 0~,
\eqno (2.29) $$
and two more equations given by permuting $ A B C $ cyclically. The equation
(2.29) also holds with $ \Psi_{21} $ replaced by $ \Psi_{22} $.

There is a duality between wave functions $ \Psi \( e^{A A'}_{~~~~i},
\psi^A_{~~i} \) $ and wave functions $ \wti \Psi \( e^{A A'}_{~~~~i}
\right . $,
$ \left .
\wti \psi^{A'}_{~~i} \) $, given by a fermionic Fourier transform [14]. The
$ S_A $ and $ \ol S_{A'} $ operators interchange r\^oles under this
transformation, and the r\^oles of $ \Psi_0 $ and $ \Psi_6,~\Psi_{21} $ and $
\Psi_{42} $, and $ \Psi_{22} $ and $ \Psi_{41} $ are interchanged. We shall
proceed by showing that $ \Psi_{22},~\Psi_{21} $ and $ \Psi_0 $ must vanish
for $ g \ne 0 $ (or $ \Lam \ne 0 $), and hence by the duality the entire
wave function must be zero.

Consider first the equation (2.29) and its permutations for $ \Psi_{21} $ and
$ \Psi_{22} $. One can check that these are equivalent to
$$ \hbar \kap^2 \( A {\pt \Psi_{21} \over \pt A} - B {\pt \Psi_{21} \over \pt
B} \) = 16 \pi^2 \( B^2 - A^2 \) \Psi_{21} \eqno (2.30) $$
and cyclic permutations. One can then integrate Eq.~(2.30) along a
characteristic $ A B = $ const., $ C = $ const., using the parametric
description $ A = w_1 e^\tau $, $ B = w_2 e^{- \tau} $, to obtain
$$ \Psi_{21} = h_1 (A B, C) \exp \[ - {8 \pi^2 \over \hbar
\kap^2} ~ \( A^2 + B^2 \) \]~. \eqno (2.31) $$
Replacing $A,B$ for $B,C$ in Eq. (2.30) gives the solution
$$ \Psi_{21} = h_2 (B C, A) \exp \[ - {8 \pi^2 \over \hbar
\kap^2} ~\( B^2 + C^2 \) \]~. \eqno (2.33) $$
Eqs.~(2.31) and (2.33) are only consistent if $ \Psi_{21} $ has the form
$$ \Psi_{21} = F (A B C) \exp \[ - {8 \pi^2 \over \hbar
\kap^2} ~\( A^2 + B^2 + C^2 \) \]~. \eqno (2.34) $$
Similarly
$$\Psi_{22} = G (A B C) \exp \[ - {8 \pi^2 \over \hbar
\kap^2} ~\( A^2 + B^2 + C^2 \) \]~. \eqno (2.35) $$

Substituting Eqs.~(2.34),(2.35) into Eq.~(2.24), one obtains
$$ \eqalignno {
&16 \pi^2 g \Psi_0
= - 2 \pi^2 \hbar (A B C)^{- 1} \( A^2 + B^2 + C^2 \)~(\exp) F \cr
&+ {3 \over 16} \hbar^2 \kap^2 (\exp) F^{\prime}
+ {2 \over 3} \( 16 \pi^2 \) \hbar (A B C)^{- 1} \( 2 A^2 - B^2 - C^2
\)~(\exp) G &(2.36) \cr } $$
and cyclically, where
$$ \exp = \exp \[ - {8 \pi^2 \over \hbar^2 \kap^2}~\( A^2
+ B^2 + C^2 \) \]~. \eqno (2.37) $$
Now $ \Psi_0 $ should be invariant under permutations of $ A, B, C $. Hence $
G = 0 $. I.e.
$$ \Psi_{22} = 0~. \eqno (2.38) $$
The equation (2.36) and its cyclic permutations, with $ \Psi_{22} = 0 $, must
be solved consistently with Eq.~(2.23) and its cyclic permutations.
Eliminating $ \Psi_0 $, one finds
$$ \eqalignno {
&{3 \hbar^3 \kap^4 \over 16 \( 16 \pi^2 g \)} F'' - {\hbar^2 \kap^2 \over 8
g}~{\( A^2 + B^2 + C^2 \) \over A B C} F' \cr
&+ 6 \pi^2 \hbar g F - {\hbar^2 \kap^2 \over 4 g}~{1 \over B^2 C^2} F +
{\hbar^2 \kap^2 \over 8 g}~{\( A^2 + B^2 + C^2 \)
\over (A B C)^2} F = 0~, &(2.39) \cr } $$
and cyclic permutations. Since $ F = F (A B C) $ is invariant under
permutations, the $ (B C)^{- 2} F $ term and its permutations imply $ F = 0 $.
Thus
$$ \Psi_{21} = 0~. \eqno (2.40) $$
Hence, using Eq.~(2.36),
$$ \Psi_0 = 0~. $$

Then we can argue using the duality mentioned earlier, to conclude that
$$ \Psi_{41} = \Psi_{42} = \Psi_6 = 0~. \eqno (2.41) $$
Hence there are no physical quantum states obeying the constraint equations
in the diagonal Bianchi-IX model. This result will be discussed further in
Sec. V.

The same conclusion can be reached for the
 case of a (non-diagonal) Bianchi type
I model. Following ref. [26], one can use
 the averaged ordering [5], with
$$ \eqalignno {
p_{A A'}^{~~~~i} \bar \psi^{A'}_{~~i} &\lrta \half \( \bar \psi^{A'}_{~~i}
p_{A A'}^{~~~~i} + p_{A A'}^{~~~~i} \bar \psi^{A'}_{~~i} \)~, \cr
\psi^A_{~~i} p_{A A'}^{~~~~i} &\lrta \half \( \psi^A_{~~i} p_{A A'}^{~~~~i} +
p_{A A'}^{~~~~i} \psi^A_{~~i} \) ~. &(2.42) \cr } $$
With this ordering, there is a certain symmetry between the operators $ S_A $
and $ \bar S_{A'} $ as viewed in the two representations $ \Psi \( e^{A
A'}_{~~~~i}, \psi^A_{~~i} \) $ and $ \wti \Psi \( e^{A A'}_{~~~~i}, \wti
\psi^{A'}_{~~i} \) $, provided one changes $ g \to - g $.
However, the final result is not an artefact of the symmetric factor ordering
(2.42) used
here. One can repeat the calculations using a general factor ordering
$$ \eqalignno {
p_{A A'}^{~~~~i} \bar \psi^{A'}_{~~i} &\lrta \( \half + s \) \bar
\psi^{A'}_{~~i} p_{A A'}^{~~~~i} + \( \half - s \) p_{A A'}^{~~~~i} \bar
\psi^{A'}_{~~i}~, \cr
\psi^A_{~~i} p_{A A'}^{~~~~i} &\lrta \( \half - s \) \psi^A_{~~i} p_{A
A'}^{~~~~i} + \( \half + s \) p_{A A'}^{~~~~i} \psi^A_{~~i}~, &(2.43) \cr
} $$
to reach the same conclusion.

Let us begin with the constraint $ \bar S_{A'} \Psi = 0 $ at order $ \psi^1
$. This gives
$$ \eqalignno {
\biggl ( - {3 \over 4} \hbar g \Psi_{2 1} & + {4 \over 3} \hbar \kap^2
h^{\half}
h_{i j} {\pt \Psi_0 \over \pt h_{i j}} - \hbar \kap^2 \Psi_0 \biggr ) n_{A A'}
\beta^A \cr
{}~& - \half e_{B B' i} n_C^{~~B'} e_{A A' j} \rga^{A B C} {\pt \Psi_0 \over
\pt h_{i
j}} = 0~, &(2.44) \cr } $$
for all $ \beta^A $ and $ \rga^{A B C} $. Take the symmetrized coefficient
of $\rga^{A B C} = \rga^{(A B C)}$ and contract it with  $ e^{A A'}_{~~~~k}
n^{B C'} e^C_{~~C'\ell} $
to get
$$ \(3 h_{i k} h_{j \ell} -  h_{k \ell} h_{i j} \) {\pt \Psi_0 \over \pt h_{i
j}} = 0~. \eqno (2.45) $$
Since the ${{\pt ~h}\over{\pt~h_{ij}}} = h~h^{ij}$, the general
solution of (2.45)
may be taken in the form
$$\Psi_0 = A~f(u),~u\equiv B~h^m, \eqno (2.46)$$
where $A, B, m$ are constants. Taking now the $\beta^A $ part of Eq.(2.44) and
using
Eq. (2.46) we get
$$ - {3 \over 4} \hbar g \Psi_{2 1}  + 4  \hbar \kap^2 A f^{'} B m h^{m+\half}
 - \hbar \kap^2 A f = 0~,\eqno (2.47)$$
where $(')$ denores derivation with respect to the $u$ variable.
Then, one gets as solutions
$$ \Psi_{2 1} = {4 \over 3} g^{-1} A \kap^2 (4 f^{'} B m h^{m+\half} - f)~.
\eqno (2.48)$$

The constraint $ S_A \Psi = 0 $ at order $ \psi^1 $ can be
shown to yield
$$ 2 g h^\half \Psi_0 \beta_A - {1 \over 4} \hbar^2 \kap^2
h^{-\half} h_{ij}
{\pt \Psi_{2 1} \over \pt h_{i j}}\beta_A
- {8 \over 3} \hbar^2 \kap^2 h^{-\half} \Psi_{2 2} \beta_A $$
$$
- {9 \over 8} \hbar^2 \kap^2  \Psi_{2 1} \beta_A
- 4 e_{A A' k} D^{B A'}_{~~~~j i} n_E^{~~C'} e_{D C'}^{~~~~j} \eps_{B C}
h^{m+\half}
{\pt \Psi_{2 2} \over \pt h_{i k}} \rga^{C D E} \eqno (2.49) $$
for all $ \beta^A $ and $ \rga^{C D E} $. From the $ \rga^{C D E} $ part of
Eq.(2.49) one gets an equation for $\Psi_{2 2}$ like Eq.(2.45) and then the
solutions $\Psi_{2 2}$ are of the form of Eq.(2.48). Now let us consider the
 $ \bar S_{A'} \Psi = 0 $ at order $ \psi^3 $. This
reads
$$ \hbar g e^B_{~~A' j} e_{B C'}^{~~~~j} n_C^{~~C'} \Psi_{41} \beta^C \rga_{D
E F} \rga^{D E F} -  \hbar \kap^2 h^{\half} {8\over 3}  n^B_{~~C'} e^{E
C'}_{~~~~k} e_{B A' j} {\pt
\Psi_{22} \over \pt h_{j k}} \beta_E \rga_{A C D} \rga^{A C D}
 +
 $$
$$ 2 \hbar \kap^2 h^{\half} n^{E}_{~C'} e^{FC'}_{~~k} e_{BA'j} \left( {\pt
\Psi_{21} \over \pt h_{j k}} \beta^A\beta_A \rga^{B}_{~FE} +
 {\pt
\Psi_{22} \over \pt h_{j k}} \rga^{B}_{~FE}\rga_{A C D} \rga^{A C D}\right)$$
$$-\half \hbar \kap^2\left(
\half n_{AA'}\rga^{ADP}\rga_{FDP}\beta^F - {4\over 3}
n^P_{~A'} \beta^{D}\rga_{FDP}\beta^F\right)
\Psi_{2 1}
$$
$$ - \half \hbar \kap^2 \Psi_{22}
\left [
\matrix {
{8 \over 3} n_{C A'} \beta_B \rga^B_{~~D A} \rga^{A C D} \cr
- 2 n^E_{~~A'} \beta_A \rga_{B D A} \rga^{A B D} \cr
+ 4 n^E_{~~A'} \rga^B_{~~D A} \rga^{A C D} \rga_{C B E} \cr
} \right ]~= 0~, \eqno (2.50) $$
for all $ \beta^A $ and $ \rga^{B C D} $.
Taking the part
$$-4 \hbar \kap^2 h^{\half} n^{E}_{~C'} e^{FC'}_{~~k} e_{BA'j}
 {\pt
\Psi_{22} \over \pt h_{j k}} \rga^{B}_{~FE}\rga_{A C D} \rga^{A C D}
 - 2 \hbar \kap^2 n^E_{~~A'} \rga^B_{~~D A} \rga^{A C D} \rga_{C B E}
\Psi_{22}=0~,$$
and substituting the solution for $\Psi_{2 2}$ of the form of Eq.(2.46)
we get that the term in $\rga^{B}_{~FE}\rga_{A C D} \rga^{A C D}$
is zero and so $\Psi_{2 2} =0$. Using this result back into Eq. (2.49)
together with Eq. (2.46) and (2.49)  we obtain the following equation
for $f$:
$$ \eqalignno {
2 g A h f & + {3 \over 4} \hbar \kap^4 g^{-1} A f - 2 \hbar \kap^4 g^{-1}
A B m h^{m + \half}  f^{'}  \cr
{}~&
-4 \hbar \kap^4 g^{-1} A B m^2 h^{m} f^{'} - 4 \hbar \kap^4 A B^2 m^2 h^{2m +
\half}
f^{''} =0~.~,& (2.51)\cr } $$
Expanding $f, f^{'}$ and $f^{''}$ in a power series with respect to
the $u$-variable and substituting back into Eq.(2.51)
one concludes that for the
constraint to be satisfied at any moment of time one needs that
$A=0$, this to say that $\Psi_0 = \Psi _{2 1} =0$.
 Using the duality mentioned earlier, we again argue to conclude that
$ \Psi_{41} = \Psi_{42} = \Psi_6 = 0.$
 Hence, there  are
{\it no} physical states in the Bianchi - I case either.

\bigbreak
\noindent
{\bf III THE $ k = + 1 $ FRIEDMANN MODEL WITH  $ \Lam $-TERM}

The $ k = + 1 $ Friedmann model without a $ \Lam $ term has been discussed
in [2,6]. There are two linearly independent physical quantum states. One is
bosonic and corresponds to the wormhole state [16], the other is at quadratic
order in fermions. The Hartle--Hawking state [17] is also   found
[15]. In the
Friedmann model with $ \Lam $ term, the coupling between the different
fermionic levels `mixes up' this pattern [4].

In the Friedmann model, the wave function has the form [6]
$$ \Psi = \Psi_0 (A) + \( \beta_C \beta^C \) \Psi_2 (A)~. \eqno (3.1) $$
As part of the Ansatz of [6], one requires $ \psi^A_{~~i} = e^{A A'}_{~~~~i}
\wti \psi_{A'} $ and $ \wti \psi^A_{~~i} = e^{A A'}_{~~~~i} \psi_A $; this is
in order that the form of the one-dimensional Ansatz should be preserved
under one-dimensional local supersymmetry, suitably modified by local
coordinate and Lorentz transformations. Thus the gravitino field is truncated
to spin $ {1 \over 2} $. Note that $ \beta^A = {3 \over 4} n^{A A'} \wti
\psi_{A'} $.

One then proceeds as in Sec.~II to derive the consequences of
the $ \ol S_{A'} \Psi = 0 $ and $ S_A \Psi = 0 $ constraints at level $
\psi^1 $, by writing down the general expression for a constraint and then
evaluating it at a Friedmann geometry. Note that it is not equivalent to
set $ A = B = C $ in Eqs.~(2.23) and (2.24); the coefficients in the
constraint equations are different. One then obtains
$$ \hbar \kap^2 {d \Psi_0 \over d A} + 4 8 \pi^2 A \Psi_0 + 1 8 \pi^2 \hbar g
A^2 \Psi_2 = 0 \eqno (3.2) $$
and
$$ \hbar^2 \kap^2 {d \Psi_2 \over d A} - 4 8 \pi^2 \hbar A \Psi_2 - 2 5 6
\pi^2 g A^2 \Psi_0 = 0~. \eqno (3.3) $$
These give second-order equations, for example
$$ A {d^2 \Psi_0 \over d A^2} - 2 {d \Psi_0 \over d A}
+ \[ - {4 8 \pi^2 \over \hbar \kap^2} A - {(48)^2 \pi^4 \over \hbar^2
\kap^4} A^3 + {9 \times 512 \pi^4 g^2 \over \hbar^2 \kap^4} A^5 \] \Psi_0 =
0~. \eqno (3.4) $$
This has a regular singular point at $ A = 0 $, with indices $ \lam = 0 $ and
3. There are two independent solutions, of the form
$$ \eqalignno {
\Psi_0 &= a_0 + a_2 A^2 + a_4 A^4 + \ldots~, \cr
\Psi_0 &= A^3 \( b_0 + b_2 A^2 + b_4 A^4 + \ldots \) ~, &(3.5) \cr } $$
convergent for all $ A $. They obey complicated recurrence relations, where
(e.g.) $ a_6 $ is related to $ a_4,~a_2 $ and $ a_0 $.

One can look for asymptotic solutions of the type $ \Psi_0 \sim \( B_0 +
\hbar B_1 + \hbar^2 B_2 + \ldots \) \exp $ \break
$ ( - I / \hbar ) $, and finds
$$ I = \pm {\pi^2 \over g^2} \( 1 - 2 g^2 A^2 \)^{3 \over 2}~, \eqno (3.6) $$
for $ 2 g^2 A^2 < 1 $.
The minus sign in $ I $ corresponds to taking the action of the classical
Riemannian solution filling in smoothly inside the three-sphere, namely a
portion of the four-sphere $ S^4 $ of constant positive curvature. This gives
the Hartle--Hawking state [17]. For $ A^2 > \( 1 / 2 g^2 \) $, the Riemannian
solution joins onto the Lorentzian solution [36]
$$ \Psi \sim \cos \left \{ \hbar^{- 1} \[ {\pi^2 \( 2 g^2 A^2 - 1 \)^{3 \over
2} \over g^2} - {\pi \over 4} \] \right \}~, \eqno (3.7) $$
which describes de Sitter space-time.

\bigbreak
\noindent
{\bf IV OTHER APPROACHES: $\sigma-$MODEL SUPERSYMMETRIC EXTENSION AND
ASHTEKAR CANONICAL QUANTIZATION}

In this section we  describe briefly other approaches to study
the quantization of cosmological models with supersymmetry, which alllows one
to extract similar conclusions.

The $\sigma-$model supersymmetric extension in quantum cosmology
has been
developed by R. Graham and collaborators [4,5,28,29]. As  is well known,
the geometrodynamics of the Bianchi models reduce, formally, to the
Hamiltonian dynamics of a particle
with coordinates $q^{\ti{\mu}}$ in a three or two dimensional
potential $V(q^{\ti{\mu}})$  [35]. In this approach, quantum models are
constructed by coupling additional fermionic degrees of freedom
to the purely gravitational ones (the minisuperspace vielbein) in such a way
that the coupled system acquires a larger symmetry, namely
local supersymmetry. The supersymmetric  extension of a particle motion
 in a potential well is treated by supersymmetric quantum mechanics [28].
The case of dynamics on a curved manifold with metric
$ds^2 = \ti{g}_{\ti{\mu}\ti{\nu}} d q^{\ti{\mu}}   d q^{\ti{\nu}}$,
where $\ti{g}_{\ti{\mu}\ti{\nu}}$ is the minisuperspace metric, has been
studied in the N=2 supersymmetric model $\sigma-$model [28]. Supersymmetry
than requires that  the potential $V(q^{\ti{\mu}})$  must be
derivable from an underlying superpotential $\Phi(q^{\ti{\mu}})$ as
$$  V(q^{\ti{\mu}}) = \half  \ti{g}_{\ti{\mu}\ti{\nu}}
  {\pt\Phi \over {\pt  q^{\ti{\mu}}}} {\pt\Phi \over{ \pt q^{\ti{\nu}}}}~.
  \eqno (4.1)       $$
Such conditions were verified for the Bianchi type I, II, VII, VIII, IX,
Kantowski-Sachs, Taub and Taub-Nut models.

It is important to note that this \susy ~extension of Hamiltonian
dynamics of a particle only leads to a N=2 \susy . By contrast,
from (1+3)  dimensional N=1 supergravity  a
dimensional reduction allows one to
obtain a (1+0)-dimensional theory with N=4 \susy [28,29]. (The extension of R.
Graham's approach to N=4
is non-trivial but such an extension has  been provided for the case
of a Bianchi type-IX without matter [29]). One may consider the N=2 \susy
{}~as a {\it sub}symmetry of the larger N=4 \susy ~ obtained from supergravity
and an attempt to clarify that connection explicitly would be interesting.
In particular [28], while in the N=2 supersymmetry ~extension it
is possible to construct
solutions in all fermionic sectors, this is different from N=4
supersymmetry
{}~minisuperspace models, in which an additional internal rotational symmetry
inherited from the Lorentz invariance of supergravity
rules out all states except those in the empty and filled fermionic sectors.
It is also curious to mention, even if not yet clear, that if the
minisuperspace
metric is Wick-rotated (i.e., the scale factor is complexified) this
leads to the same restrictions of physical states as the requirement
of all the symmetries included in the N=4 \susy ~models.

The application of the $\sigma-$model supersymmetry
{}~extension programme to a general
non-diagonal Bianchi type-IX model with a cosmological constant term is given
in refs. [5,28]. The complete Hamiltonian and the classical constraints of the
 model are then derived. The system is quantized {\it \` a la} Dirac,
replacing brackets by commutators or anti-commutators and the canonical
momenta by appropriatr derivatives with respect to the canonical coordinates.
However, this \susy ~framework was then applied to the particular case of a
closed Friedmann-Robertson-Walker case.
The wave function of the Universe can be written as
$$ |\Psi\rangle = (\Psi_0 + \Psi_a c_a^{\dagger} +
\Psi_{a\Lambda} c_{\Lambda}^{\dagger} c_a^{\dagger}
 + \Psi_{\Lambda} c_{\Lambda}^{\dagger}
 )|0\rangle~,
\eqno (4.2)      $$
where $c_a^{\dagger},c_{\Lambda}^{\dagger}$ stand for a set of creation
operators that replace the fermionic partners in a Fock  state representation;
$|0\rangle $.

 From  the N=2 \susy ~constraints R. Graham obtains that the admissable
solution
is of the form
$$ \Psi_0 = \Psi_{a\Lambda} =0 ~,\eqno(4.3)$$
$$\Psi_{a(\Lambda)}^{WBK} = \sqrt{(-)1 + \left[1-{{2\Lambda}\over{9\pi^2}}6\pi
a^2\right]^{-\half} }
\exp\left(-\half \int_0^{6\pi a^2} du
\sqrt{ 1-{{2\Lambda}\over{9\pi^2}}u}\right) ~,\eqno(4.4)$$
which in the limit $a\rightarrow 0$ gives
$\Psi_a \rightarrow 0;~\Psi_{\Lambda}\rightarrow \exp(-\Phi)$
where $\Phi$ is the superpotential previously described and
$\Phi = 3\pi a^2~.$
For $6\pi a^2 > \left({{2\Lambda}\over{9\pi^2}}\right)^{-1}$ the exponent in
is oscillatory.
These results  point tothe wave function as
being in the
the initial state  found by Vilenkin [37],
apart from the appearance of additional Grassmann
variables.

Now, let us make  some brief comments about
canonical quantization of supergravity using Ashtekar variables.
Recently Ashtekar has presented a new formulation of  Einstein gravity. One of
the remarkable
features of this formalism is that the constraints of gravity are simple
polynomials of the canonical
variables. So the constraints are more manageable than
the ones expressed in the  metric representation.
One may hope that this feature persists
 in the canonical quantization of N = 1 supergravity. This
turns out to be true: the supersymmetry
constraints are again simple polynomials of the canonical
variables.

The Ashtekar formalism of the N = 1 supergravity has been formulated by
Jacobson [31]. Capovilla and
Guven [38] have successfully carried out the quantization of all the Bianchi
type A models using this
formalism. They obtained similar results to  D'Eath [9].  Recently Capovilla
and Obregon [33], Sano and
Shiraishi [32] have also studied the quantization of the N = 1 supergravity
with a cosmological constant.
In [32]
they found a semi-classical solution in the
full theory. It has the form of exponential of the N =1
supersymmetric extension of the Chern-Simons functional.
They applied this semi-classical wavefuction to the FRW universe using
WKB approximation. The general line element is
$$ ds^{2} = -d \tau^{2} + {\sigma \over 8} d^{2} \Omega~, \eqno (4.5)$$
They have considered four cases:
$ \tau :{\rm  real},~ \sigma < 0;~ $
$  \tau : {\rm   imaginary},~\sigma > 0;~ $
$ \tau : {\rm  real},~ \sigma > 0 ;~$
$ \tau : {\rm  imaginary },~\sigma < 0. $~
They try to find the classical solution of $\sigma$ under these 4 cases.
In the first case, the universe has the form of an Euclidean hyperbola.
In case 2, the universe is a 4-sphere.
In case 3, it  is an open universe.
In the last case, there is no solution.

The reality condition on the gravitino  is the Majorana
condition. Their solution does not satisfy this condition in general. To
obtain the real solution, they must transform the solutions by the
transformations corresponding to the symmetries in the theory. They did not
solve this problem in [32] and will consider it on another occasion.

 In [33],  the quantization of  the
class A Bianchi Models was studied.
 In the following, we describe briefly the work of Capovilla and Obreg\'on.

The canonical variables   are
$$ \tilde{\sigma}^{i}_{AB}, A_{i}^{~AB}, \psi_{iA}, \tilde{\pi}^{iA}
{}~, \eqno (4.6) $$
where $ A_{i}^{~AB}$ is a self-dual connection,
 $ \tilde{\sigma}^{i}_{AB}$ is a  densitized SU(2) soldering form where
$$ \tilde{\sigma}^{i}_{AB} = i h^\half  n^{AA'} e^{iB}_{~~A'}~, \eqno (4.7) $$
and  $det(q) q^{ij} = \tilde{\sigma}^{iAB}  \tilde{\sigma}^{j}_{~AB}$.
Here $\tilde{\pi}^{iA}$ is the  momentum conjugate to $ \psi_{iA}$.

In the class A Bianchi Models, a triad of basis  vectors satifsy
$$  [ X_{a} , X_{b} ]^{i} = C_{ab}^{~~~c}  X^{i}_{c}~, \eqno (4.8)$$
where $ C_{ab}^{~~~c} = \epsilon_{abc} M^{cd}$ denote the structure constant
and $M^{ab} = M^{ba}$.

The Jacobson phase space variables may be expanded with respect to the
triad vectors and their duals $ \chi_{i}^{a} $
$$ A_{iA}^{~~B} = A_{aA}^{~~B} \chi_{i}^{a}~, \eqno (4.9a)$$
$$ \tilde{\sigma}^{iAB} = det(\chi) \sigma^{aAB} X^{i}_{a}~, \eqno (4.9b) $$
$$ \psi_{i}^{~A} = \psi_{a}^{~A} \chi_{i}^{a}~, \eqno (4.9c) $$
$$ \tilde{\pi}^{iA} = det(\chi) \pi^{aA} X^{i}_{a}~, \eqno (4.9d) $$
The fundamental Poisson brackets are given by
$$ \{\sigma^{aAB}, A_{bCD}\} = {i\over\sqrt{2}}
\delta^{b}_{a}\delta_{(C}^{~~A} \delta_{D)}^{~~B}~, \eqno (4.10) $$
$$ \{\pi^{aA} , \psi_{bB}\} = {i\over\sqrt{2}}  \delta^{a}_{b} \delta_{B}^{~A}
{}~, \eqno (4.11)  $$

Because of the use of the self-dual connection as a field variable, we
need to impose the {\it reality condition} [39] for the construction   of the
inner product of the physical wave function. However,  we will  not need
it  in here because one  finds that the only physical
 wave function is the trivial one.
In the quantum theory we get from the triad representation
$$ A_{a}^{~AB} = {1\over\sqrt{2}}  { {\pt }\over {\pt \sig^{a}_{~AB}}} ~, \eqno
(4.12)  $$
$$ \pi^{a}_{~A} = {1\over\sqrt{2}} { {\pt }\over {\pt \psi_{a}^{~A}}} ~, \eqno
(4.13)     $$
Using this ansatz, one obtains
$$ S^{A} = \half   {{\pt^{2}}\over{\pt  \sig^{a}_{AB} \pt   \psi_{aA}^{~~~B}
}}  + i2\sqrt{2}m \left(\sigma^{a} \psi_{a}\right)^{A}~, \eqno (4.14)   $$
$$ \bar{S}^{A} = - \half \epsilon_{abd} M^{dc} (\sigma^{a}\sigma^{b}
\psi_{c})^{A}
+ {1\over\sqrt{2}}\left(\sigma^{[a}\sigma^{b]}{{\pt
\psi_{b}}\over{\pt\sig^{a}}}\right)^{A}
 -i2mh^{-\half}\left(\sigma_{a}{\pt\over{\pt\psi_{aA}}}\right)^{A}~,
\eqno (4.15)  $$
where $\Lambda = -4m^{2}$

The last terms in  $S^{A},\bar{S}^{A}$ are the contributions from the
cosmological constant.
 The supersymmetry
constraints  here are simpler than those  in
the  metric formulation. Using the same decompositon of the gravitino field and
the same ansatz of the wavefunction,  one again  obtains a set of equations.
The only solution which
 satisfies this set of equations is the trivial solution: there are no physical
states. The simple polynomial form of the  constraints suggests
that the  study of the
 full theory  with a cosmological constant term might
  be easier in this formalism.

\bigbreak
\noindent
{\bf V CONCLUSION}

There are no physical quantum states for $ N  = 1 $
supergravity with a $ \Lam $-term in the  Bianchi models.  The physical
states found in Sec.~III for the $ k = + 1 $ Friedmann model, where the
degrees of freedom carried by the gravitino field are $ \beta_A $, disappear
when the further fermionic degrees of freedom $ \rga_{A B C} $ of the
Bianchi-IX model are included.

One could also study this from the point of view of perturbation theory about
the $ k = + 1 $ Friedmann model. As well as the usual gravitational harmonics
[40], gravitino harmonics can be used [41]. For example, the Bianchi-IX model
with radii $ A, B, C $ close together describes a particular type of
`gravitational wave' distortion of the Friedmann model; similarly for the $
\rga_{A B C} $ of the Bianchi-IX model, which describes a particular
`gravitino wave' distortion. Quite generally, in perturbation theory [40,42]
one expects to find a wave function which is a product of the background wave
function $ \Psi (A) $ times an infinite product of wave functions $ \psi_n
$
(perturbations) where $ n $ labels the harmonics. And one further expects that
the
perturbation wave function corresponding to the Bianchi-IX modes must be zero,
by a perturbative version of the argument of Sec.~II. [It will be
interesting to investigate this.] Hence the complete perturbative wave
function should be zero; then physical states would be forbidden for a generic
model of the gravitational and gravitino fields with $ \Lam $-term. This
suggests that the full theory of $ N = 1 $ supergravity with a
non-zero $ \Lam $-term should have {\it no} physical states.
\bigbreak
\noindent
{\bf ACKNOWLEDGEMENTS}
The authors are very grateful to P.K. Townsend, H. Kodama for
discussions and helpful suggestions.
A.D.Y.C. thanks the Croucher Foundation of Hong Kong for financial support.
P.R.L.V.M. is very grateful to Prof. V.N. Melnikov and to the Organizing
Committee of the International School-Seminar
"Multidimensional Gravity and Cosmology" for all their
kind assistance.
 P.R.L.V.M. also gratefully
acknowledges  the support of a Human Capital and Mobility
grant from the European Union (Program ERB4001GT930714).
\vskip .2 true in
\noindent
{\bf REFERENCES}

{\rm

\advance\leftskip by 4em
\parindent = -4em

[1] A. Maci\'as, O. Obreg\'on and M.P. Ryan, Class.~Quantum Grav.~{\bf 4},
1477 (1987).

[2] P.D. D'Eath and D.I. Hughes, Phys.~Lett.~{\bf 214}B, 498 (1988).

[3] R. Graham, Phys.~Rev.~Lett.~{\bf 67}, 1381 (1991).

[4] R. Graham, Phys.~Lett.~{\bf 277}B, 393 (1992).

[5] R. Graham and J. Bene, Phys.~Lett.~{\bf 302}B, 183 (1993).

[6] P.D. D'Eath and D.I. Hughes, Nucl.~Phys.~B {\bf 378}, 381 (1992).

[7] L.J. Alty, P.D. D'Eath and H.F. Dowker, Phys.~Rev.~D {\bf 46}, 4402
(1992).

[8] P.D. D'Eath, S.W. Hawking and O. Obreg\'on, Phys.~Lett.~{\bf 300}B, 44
(1993).

[9] P.D. D'Eath, Phys.~Rev.~D {\bf 48}, 713 (1993).

[10] M. Asano, M. Tanimoto and N. Yoshino, Phys.~Lett.~{\bf 314}B, 303 (1993).

\break

[11] A.D.Y. Cheng, P.D. D'Eath and
P.R.L.V. Moniz, {\it  Canonical Formulation  of N=1
Supergravity with Supermatter}~DAMTP R94/13, submitted to
Physical Review D;{\it  Quantization of a Friedmann Model with Supermatter
in N=1 Supergravity}~DAMTP R94/21, gr-qc 9406048.

[12] P. D. D'Eath and P. V. Moniz, {\it  Supersymmetric Quantum  Cosmology},
work in preparation.

[13] C. Teitelboim, Phys.~Rev.~Lett.~{\bf 38}, 1106 (1977).

[14] P.D. D'Eath, Phys.~Rev.~D {\bf 29}, 2199 (1984).

[15] R. Graham and H. Luckock, Sidney Univ. Maths. Report. 93-06,
gr-qc 9311004.

[16] S.W. Hawking and D.N. Page, Phys.~Rev.~D {\bf 42}, 2655 (1990).

[17] J.B. Hartle and S.W. Hawking, Phys.~Rev.~D {\bf 28}, 2960 (1983).

[18] P.D. D'Eath,  Phys.~Lett.~B {\bf 321}, 368 (1994).

[19] P.K. Townsend, Phys, Rev.~D {\bf 15}, 2802 (1977).

[20] P. van Nieuwenhuizen, Phys. Rep. {\bf 68}, 189 (1981).

[21] A. Das and D.Z. Freedman, Nucl. Phys. B{\bf 120}, 271 (1979).

[22] E.S. Fradkin and M.A. Vasiliev, Lebedev Institute
preprint No.197 (1976).

[23] D.Z. Freedman, Phys. Rev. Lett. {\bf 38}, 105 (1977).

[24] S. Ferrara, J. Scherk and B. Zumino, Phys. Lett. {\bf 66}B, 35 (1977).

[25] S. Deser and B. Zumino, CERN report 1977

[26] P.D. D'Eath, Phys. Lett. B{\bf 320}, 20 (1994).

[27]  A.D.Y. Cheng, P.D. D'Eath and
P.R.L.V. Moniz, Phys. Rev. D{\bf 49} (1994) 5246.

[28] R. Graham, Phys. Rev. D{\bf 48}, 1602 (1993).

[29] R. Graham and J. Bene,  Phys. Rev. D{\bf 49}, p.799 (1994).

[30] H. Kodama, Phys. Rev. D{\bf 42}, 2548 (1990)

[31] T. Jacobson, Class. Quantum Grav. {\bf 5}, 923, (1988).

[32]  T. Sano and J. Shiraishi, Nucl. Phys. B{\bf 410}, 423, (1993).

[33] R. Capovilla and O. Obregon, CIEA-GR-9402, gr-qc:9402043.

[34] H.-J. Matschull, DESY 94-037, gr-qc:9403034.

[35] M.P. Ryan and L.L. Shepley, {\it  Homogeneous Relativistic Cosmologies}
(Princeton University Press, Princeton, 1975).

[36] J.B. Hartle, in {\it High Energy Physics 1985}, ed. M.J. Bowick and F.
G\"ursey (World Scientific, Singapore, 1986).

[37] A. Vilenkin, Phys. Rev. D{\bf 30}, 509, (1984);
Nucl. Phys. B{\bf 252}, 141 (1985).

[38] R Capovilla and J Guven, CIEA-GR-9401, gr-qc:  9402025.

[39] A. Ashtekar,{\it Lectures on Non-Perturbative Canonical Gravity},
Singapore, Singapore: World Scientific (1991) (Advanced series in
astrophysics and cosmology, 6).

[40] J.J. Halliwell and S.W. Hawking, Phys.~Rev.~D {\bf 31}, 1777 (1985).

[41] D.I. Hughes, Ph.D.~thesis, University of Cambridge (1990), unpublished.

[42] P.D. D'Eath and J.J. Halliwell, Phys.~Rev.~D {\bf 35}, 1100 (1987).

\ \ \ }
\bye